\documentstyle[11pt,a4]{article}
\begin{document}
\begin{center}
{\bf Exact Lema\^{\i}tre-Tolman-Bondi solutions for matter-radiation
decoupling}
\end{center}

\begin{center}
Roberto  A. Sussman\\
{\small Instituto de Ciencias Nucleares UNAM, Apartado Postal 70-458,\\
M\'exico DF, 04510, M\'exico}
\end{center}

\begin{center}
Diego Pav\'{o}n\\
{\small Departamento de F\'{\i}sica, Universidad Aut\'onoma de Barcelona,\\
08193 Bellaterra, Spain}
\end{center}
\  \\
\  \\
\begin{abstract}
A new class of exact solutions of Einstein's equations is derived
providing a physically plausible hydrodynamic description of
cosmological matter in the radiation era of the Universe expansion.
The solutions are characterized by the LTB metric with dissipative
fluid sources, subjected to the following conditions: (i) the
nonequilibrium state variables satisfy the equations of state
of a mixture of relativistic and non-relativistic  ideal gases,
where the internal energy of the latter has been neglected, (ii)
the particle numbers of the mixture components are independently
conserved, (iii) the viscous stress is consistent with the transport
equation and entropy balance law of extended  irreversible
thermodynamics with the coefficient of shear viscosity provided
by kinetic theory. The Jeans mass at decoupling is of the same order
of magnitude as that of  baryon dominated models
(i.e. $M_{J} \approx 10^{16} M_{\odot}$).
\end{abstract}
\[\]
\[\]
\section{Introduction}
The standard approaches to the radiative era of cosmic evolution (the
period from the end of cosmic nucleosynthesis to the decoupling of
matter and radiation) resort to a FLRW space-time background
while the sources of the gravitational field are described either by
equilibrium kinetic theory \cite{kolb}, gauge invariant perturbations
\cite{borner}, or some hydrodynamical model \cite{coley}, which,
in general, fail to incorporate a physically plausible description of
the matter-radiation interaction since they assume thermodynamical
equilibrium throughout. Here we consider the aforesaid sources in
the temperature range $ 10^{6} \mbox{K} \leq T \leq 10^{3}\mbox{K}$,
as a nonequlibrium mixture of a non-relativistic fluid (matter) and
a extreme relativistic one (radiation); their mutual interaction is
modeled by a dissipative shear-stress tensor, and the background
space-time is described by the Lema\^{\i}tre-Tolman-Bondi (LTB for
short) metric \cite{sussman2}. (The interest for inhomogeneous metrics
has revived in the wake of the work of Mustapha {\it et. al.}
\cite{musta} who show that given any spherically symmetric
geometry and any set of observations, evolution functions for the
observed sources can be found that will make the model in general
compatible with observation).  The stress-energy tensor
reads
\[
T^{ab}= \rho u^au^b+ph^{ab}+\Pi^{ab} \qquad (h^{ab}= c^{-2}u^a
u^b+g^{ab}) \, ,
\]
where is assumed that the matter fluid is just ``dust" and therefore
the hydrostatic pressure is furnished by the radiation only, i.e.
\[
\rho\approx mc^2n^{(m)}+3n^{(r)}k_{_B}T, \qquad
p\approx n^{(r)}k_{_B}T
\]
\noindent
In its turn the shear dissipative pressure tensor is governed by
\[
\tau \dot\Pi_{cd}\,h^c_ah^d_b+\Pi_{ab}\left[
1+{1\over2}T\eta\left({{\tau}\over{T\eta}}  \,
u^c\right)_{;c}\right]+ 2\eta\,\sigma_{ab}= 0, \quad \mbox {with}
\quad
\eta_{_{(rg)}}= {4\over{5}}p^{(r)}\tau \, ,
\]
\noindent
this expression is compatible with relativistic causality and stability, and
is supported by kinetic theory, statistical fluctuation theory, and
information theory \cite{israel}. The entropy per particle takes
the form
\[
s= s^{(e)}+\frac{\alpha}{nT}\,\Pi_{ab}\Pi^{ab},\quad \Rightarrow \quad
(snu^a)_{;a}\geq 0,
\]
\noindent with
\[
\alpha_{_{(rg)}}= -
{\tau\over{2\eta_{_{(rg)}}}}= -\frac{5}{8p^{(r)}} \, ,
\]
and $\tau$ is the relaxation time for the shear-stress. Further,
since the number of particles of each fluid is independently
conserved, we have
$ (n^{(m)}u^a)_{;a} = (n^{(r)}u^a)_{;a} = 0 $.

\section{Exact Solution and Density Contrasts}
The LTB metric can be written as
\[
ds^2= -c^2dt^2+{Y'^2\over{1-F}}\,dr^2 +Y^2\left(d\theta^2+\sin^2
(\theta) d\phi^2
\right)\, , \quad Y = Y(t,r)\, , \quad F = F(r).
\]
\noindent
From this metric element the expansion and shear read
\[
\Theta= {\dot Y'\over{Y'}}+{{2\dot Y}\over Y} \, , \quad
\sigma\equiv{1\over 3}\left({\dot Y\over Y}-{\dot Y'\over Y'} \right)
\, , \quad
\sigma^a\,_b= {\bf{\hbox{diag}}}\left[0,-2\sigma,\sigma,\sigma \right],
\]
\noindent
respectively;
and the  most general form for the shear-stress reduces to
$\Pi^a\,_b= $
\newline
${\bf{\hbox{diag}}}\left[0,-2P,P,P \right]$,  with $P = P(t,r) $.
The non-trivial Einstein's field equations are
\[
\kappa \rho=-\frac{\left[Y\left( \dot Y^2+Fc^2\right)
\right]{}'}{Y^2Y'}=-G^t\,_t
\]
\[
\kappa p=-\frac{\left[Y\left( \dot Y^2+Fc^2\right) +2Y^2\ddot
Y\right]{}'}{3Y^2Y'}=\frac{1}{3}\left(2G^\theta\,_\theta+G^r\,_r
\right) \, ,
\]
\[
\kappa P=\frac{Y}{6Y'}\left[\frac{Y\left( \dot Y^2+Fc^2\right) +2Y^2\ddot
Y}{Y^3} \right]{}'= \frac{1}{3}\left(G^\theta\,_\theta-G^r\,_r
\right),
\]
\noindent
where the prime and overdot mean partial derivatives with respect
to $r$ and $t$, respectively.
In the case of flat spatial sections $(F = 0)$, integration  of
these equations yields the following exact solution \cite{sussman2}
\[
\frac{3}{2}\sqrt{\mu} \,(t-t_i)=
\sqrt{y+\epsilon}\left(y-2\epsilon
\right)-
\sqrt{1+\epsilon}\left(1-2\epsilon\right) ,
\]
\noindent
where we have used the definitions
\[
\mu\equiv  \frac{\kappa M}{Y_i^3},\qquad \epsilon\equiv
\frac{W}{M},\qquad
y\equiv\frac{Y}{Y_i} \, ,
\]
\noindent
with
\[
M= \int{\rho_i^{(m)}Y_i^2Y'_idr},\qquad \rho_i^{(m)}\equiv mc^2
n_i^{^{(m)}} \, , \qquad
W= \int{\rho_i^{(r)}Y_i^2Y'_idr},\qquad \rho_i^{(r)}\equiv
3n_i^{^{(r)}}k_{_B}T_{i} \, ,
\]
\noindent
so that $\rho_i^{(m)},\,\rho_i^{(r)} $ define the initial
densities of the non-relativistic and  relativistic
components of the mixture, respectively.

At this point it is expedient to introduce the density contrast parameters
\[
\rho_i^{(m)}=\left\langle
\rho_i^{(m)}\right\rangle\left[1+\Delta_i^{^{(m)}}\right], \qquad
\rho_i^{(r)}=\left\langle
\rho_i^{(r)}\right\rangle\left[1+\Delta_i^{^{(r)}}\right],
\]
\noindent
as well as the  ancillary functions
\[
\Gamma\equiv \frac{Y'/Y}{Y'_i/Y_i}\qquad \Psi\equiv 1+\frac{\left(1-\Gamma
\right)}{3(1+\Delta_i^{^{(r)}})} ,\qquad
\Phi\equiv
1+\frac{\left(1-4\Gamma
\right)}{3(1+\Delta_i^{^{(r)}})} ,
\]
the latter are linked to the density contrasts by
\[
\Gamma= 1+3A\Delta_i^{^{(m)}}+3B\Delta_i^{^{(r)}} \, ,
\]
\[
\Psi= 1-\frac{A\Delta_i^{^{(m)}}
+B\Delta_i^{^{(r)}}}{1+\Delta_i^{^{(r)}}} \, ,
\]
\[
\Phi= \frac{-4A\Delta_i^{^{(m)}}
+\left(1-4B\right)\Delta_i^{^{(r)}}}{1+ \Delta_i^{^{(r)}}} \, ,
\]
where the quantities $A $ and $B$ are known functions of $y$.
Given a set of initial conditions specified by
$\epsilon,\,\Delta_i^{^{(m)}},\,\Delta_i^{^{(r)}}$, on some initial
hypersurface, the forms of the state and geometric variables
as functions of $y$ and the chosen initial conditions are fully
determined. However, for the solutions to be physically
meaningful they must comply with the following set of physical
restrictions,
\[
\dot s= \frac{15 k_{_B}}{16\tau}
\left(\frac{\Phi} {\Psi}\right)^2 > 0, \, \quad
\Gamma > 0, \, \quad \Psi > 0, \, \quad \sigma \Phi < 0,
\]
\[
\dot \tau>0,\qquad {\ddot s\over\dot
s}= {{2\sigma\Gamma}\over{3\Psi\Phi}}\,{{\left\langle
\rho_i^{(r)}\right\rangle}\over
{\rho_i^{(r)}}}\left[1+{{\left\langle \rho_i^{(r)}\right\rangle}\over
{3\rho_i^{(r)}}}\right]-{\dot\tau\over{\tau}} < 0,
\]
\[
\tau < 3/\Theta \quad \mbox{before decoupling}, \qquad
\tau > 3/\Theta \quad \mbox{after decoupling}.
\]

\section{Initial Perturbations and Jeans Mass}
Assuming that the perturbations on the intial hypersurface
are non-adiabatic (i.e.
$\Delta_{i}^{(s)} \equiv \frac{3}{4}\Delta_{i}^{(r)}-
\Delta_{i}^{(m)}\neq 0 $) and that $T_{i} \approx 10^{6}$ Kelvin,
the set of values associated to the matter-radiation
decoupling is obtained from solving numerically the equation \cite{sussman2}
\[
T_{_D}\approx 4\times 10^3
=\frac{10^6}{y_{_D}}\Psi(y_{_D},\Delta_i^{(s)},\Delta_i^{^{(r)}}) \, ,
\qquad (T = \frac{T_{i}}{y} \Psi),
\]
\noindent
which results in $ y_D \approx 10^{2.4}$.\\

The temperature anisotropy of the CMR ($[\delta T/T]_{_D}\approx10^{-5}$)
constrains the maximal values of
$\Delta_i^{^{(m)}},\,\Delta_i^{^{(r)}}$ to about $10^{-4}$ and
the corresponding variation range of $|\Delta_i^{(s)}|$ is
$|\Delta_i^{(s)}|<10^{-8}$, and so compatibility
with acceptable values of $T_{_D}$ and the CMR anisotropy implies
$|\Delta_i^{(s)}|\approx |\Delta_i^{^{(r)}}|^2\ll |\Delta_i^{^{(r)}}|$.\\

For non-adiabatic perturbations the Jeans mass is given by
\[
M_{_J}=\frac{4\pi}{3}mn^{(m)}\left[\frac{8\pi^2
C_s^2}{\kappa(\rho+p)}\right]^{3/2}=
 \frac{4\pi}{3}\frac{c^4\chi_0\Gamma^{1/2}}{\sqrt{\rho_i^{(r)}}}
\left[\frac{\pi y^2\Psi}{3G\left(\Psi+\textstyle{{3} \over
{4}}\chi_iy\right)^2}\right]^{3/2},
\]
\noindent
where
\[
C_s^2=\frac{c^2}{3}\left[1+\frac{3\rho^{(m)}}{4\rho^{(r)}}\right]^{-1},
\qquad \chi_i=\rho_i^{m}/\rho_i^{r}.
\]
\noindent
Assuming $y=y_{D}\approx 10^{2.4}$, $\epsilon\approx
1/\chi_i\approx
10^3$ and $\rho_i^{(r)}\approx a_{_B}T_i^4\approx 7.5\times
10^9\,\hbox{ergs}/\hbox{cm}^3$, yields $M_{_J}\approx 10^{49}\hbox{gm}$, or
approximately
$10^{16}$ solar masses. This value coincides with the Jeans mass obtained
for baryon dominated perturbative models as decoupling is approached in the
radiative era.

\section{Conclusions and Outlook}
The model presented here is a self consistent
hydrodynamical approach to matter-radiation mixtures that:
(a) is based on inhomogeneous exact solutions of
Einstein's field equations, and (b) is thermodynamically consistent.
It has, however, the limitations of not incorporating heat currents
(the LTB models do not allow for this\cite{bondi}) nor bulk dissipative
stresses.
Nonetheless, we believe this model may become a useful theoretical
tool in the study of cosmological matter sources, providing a
needed alternative and complement to the usual perturbative or
numerical approaches. \\

The solutions have an enormous potential as models in applications of
astrophysical and cosmological interest:\\

\noindent
\underline{ Structure formation in the acoustic phase}.
There is a large body of literature
on the study of acoustic perturbations in relation to the Jeans mass of
surviving cosmological condensations. Equations of state
of the type ``dust plus radiation" are oftenly
suggested in this context \cite{borner}.
Since practically all work on this topic has been carried on with
perturbations on a FLRW background, the exact solutions presented
here may be viewed as an alternative treatment for this problem.\\

\noindent
\underline{Comparison with perturbation theory}.
Our model is based on exact solutions of Einstein's field equations,
but their initial conditions and evolution can be adapted for a
description of ``exact spherical perturbations'' on a FLRW
background. It would be extremely interesting, not only to
compare the results of this approach with those of a
perturbative treatment, but to provide a physically plausible
theoretical framework to examine carefully how much information
is lost in the non-linear regime that falls beyond the
scope of linear perturbations. We have considered only
the case $F=0$, thus restricting the evolution to the
``growing mode'' since all fluid layers expand monotonously.
The study of the more general case, where  $F(r)$  is an arbitrary
function  that could change sign, would allow a comparison with
perturbations that include also a ``decaying mode'' related to
condensation and collapse of cosmological inhomogeneities.\\

\noindent
\underline{Inhomogeneity and irreversibility in primordial
density perturbations}.
The initial conditions of the models with
$\Delta_i^{(s)}\ne 0$ are set for a hypersurface with temperature
$T_i\approx 10^6$K. These initial conditions can be considered the
end product of processes characteristic of previous
cosmological history, and so the estimated value
$|\Delta_i^{(s)}|\approx 10^{-8}$, related to the
spatial variation of  photon entropy  fluctuations, can be used as a
constraint on the effects of inhomogeneity on primordial entropy
fluctuations that might be predicted by inflationary models at
earlier cosmic time. Also, the deviation from equilibrium in
the initial hypersurface (proportional to
$[\Delta_i^{^{(r)}}]^2\approx |\Delta_i^{(s)}|$) might be
helpful to understand the irreversibility associated with the
physical processes involved in the generation of primordial
perturbations \cite{linde}.

\section*{Acknowledgments}
This work has work has been partially supported by the Spanish Ministry
of Education and the National University of Mexico (UNAM), under
grants PB94-0718 and DGAPA-IN-122498, respectively.

\end{document}